\documentclass{article}
\usepackage{spconf}
\usepackage{amsmath,graphicx}
\usepackage{color, colortbl}
\definecolor{LightCyan}{rgb}{0.88,1,1}

\usepackage{cellspace}
\usepackage{makecell}

\usepackage{hyperref}
\hypersetup{
    colorlinks=true,
    linkcolor=blue,
    filecolor=magenta,      
    urlcolor=cyan,
    pdftitle={Overleaf Example},
    pdfpagemode=FullScreen,
    }

\usepackage{dblfloatfix}
\usepackage{enumitem}
\setlist{nosep, leftmargin=14pt}

\usepackage{mwe} % to get dummy images

\newcommand{\comment}[1]{}

\title{Physics-Informed autoencoder for DSC-MRI Perfusion post-processing: application to glioma grading}
\name{
\begin{tabular}{c}
Pierre Fayolle$^{1,2,5}$ \qquad 
Alexandre Bône$^1$ \qquad 
Noëlie Debs$^1$ \qquad 
Mathieu Naudin$^{2,3,5}$\\[0.3em]
\textit{Pascal Bourdon}$^{4,5}$ \qquad  
\textit{Rémy Guillevin}$^{2,3,5}$ \qquad  
\textit{David Helbert}$^{4,5}$
\end{tabular}
}

\address{
$^1$ Guerbet Research, Villepinte, France\\
$^2$ LMA, Université de Poitiers, France \\
$^3$ CHU de Poitiers, Poitiers, France\\
$^4$ XLIM, Université de Poitiers, France\\
$^5$ I3M, Common Laboratory CNRS-Siemens, France
}

\begin{document}
%\ninept
%
\maketitle
\begin{abstract}
DSC-MRI perfusion is a medical imaging technique for diagnosing and prognosing brain tumors and strokes. 
Its analysis relies on mathematical deconvolution, but noise or motion artifacts in a clinical environment can disrupt this process, leading to incorrect estimate of perfusion parameters. 
Although deep learning approaches have shown promising results, their calibration typically rely on third-party deconvolution algorithms to generate reference outputs and are bound to reproduce their limitations. 

To adress this problem, we propose a physics-informed autoencoder that leverages an analytical model to decode the perfusion parameters and guide the learning of the encoding network.
This autoencoder is trained in a self-supervised fashion without any third-party software and its performance is evaluated on a database with glioma patients.
Our method shows reliable results for glioma grading in accordance with other well-known deconvolution algorithms despite a lower computation time.
It also achieved competitive performance even in the presence of high noise which is critical in a medical environment.

\end{abstract}

\section{Introduction}
\label{sec:intro}
Dynamic Susceptibility Contrast Magnetic Resonance Imaging (DSC-MRI) perfusion is an MRI modality that involves the injection of a contrast agent that causes changes in magnetic susceptibility signals over time.
These modifications can be quantified to generate perfusion maps, which are essential for radiologists to accurately diagnose brain tumors or strokes. 

The perfusion parameters are typically obtained by deconvolution of the DSC signals with a reference signal, called arterial input function (AIF), measured in the main arteries irriguating the brain. 
Characteristics of the resulting tissue response function (TRF) are then derived to define perfusion maps such as the cerebral blood flow (CBF) and the mean transit time (MTT).
To solve this ill-posed problem, various methods have been published relying on Singular Value Decomposition (SVD)~\cite{ostergaard1996high, wu2003tracer}.
However, studies have found that these methods tend to underestimate CBF and may introduce non-physiological oscillations in TRF, even when regularization terms are applied~\cite{mehndiratta2013control}.

\comment{
Deep learning approaches have recently been proposed as an alternative, aiming to automatically generate perfusion maps by learning from third-party deconvolution algorithms as reference~\cite{ho2016temporal, kossen2023image, talebi2024deep}.
Asaduddin \textit{et al.,} were the firsts to propose in~\cite{asaduddin2024spinned} a deep learning approach that did not rely on third-party software to define reference labels. 
Instead, they used simulations to solve the deconvolution using a physics-informed neural network.
This approach outperformed other deconvolution algorithms, even with high noise images.
Nonetheless, their simulation block generates concentration curves that can be far from \textit{in vivo} data.
}

Deep learning approaches have recently been proposed as an alternative, aiming to automatically generate perfusion maps by learning from third-party deconvolution algorithms as reference~\cite{ho2016temporal, kossen2023image, talebi2024deep}.
More recently, new methods that did not rely on third-party softwares to define reference labels have been published~\cite{asaduddin2024spinned, rotkopf2024physics}.
Instead, they trained a physics-informed neural network with simulated data to solve the deconvolution.
This approach outperformed other deconvolution algorithms, even with high noise images.
Nonetheless, the simulations generate concentration curves that can be far from \textit{in vivo} data.

In this paper, we propose a physics-informed autoencoder (PHAE) trained with \textit{in vivo} data, that does not require any ground truth to perform the deconvolution and generate the perfusion maps with a high robustness to noise and a low computational cost.
Distinguishing Low Grade Glioma (LGG) from High Grade Glioma (HGG) was used as a metric to evaluate the performance of the proposed method in comparison with standard deconvolution algorithms. %such as oscillation-index SVD (oSVD)~\cite{wu2003tracer} and Tikhonov regularization~\cite{calamante2003quantification}.

\section{Methods}

\subsection{Data}
\subsubsection{Public dataset}\vspace{-0.25cm}
DSC-MRI sequences from 49 patients with glioma were collected from the public QIN-BRAIN-DSC-MRI dataset~\cite{schmainda2016glioma}. 
Among these subjects, 13 were histologically diagnosed with low-grade glioma (LGG) and 36 with high-grade glioma (HGG).
Manually-defined tumor segmentation maps, normal-appearing white matter and arterial voxels were also retrieved.
Arterial signals were then averaged to derive the patient-specific AIF. % expliquer que les segmentations sont faites sur les IRM T1 ? 
% Imaging sequences were acquired with either 1.5T or 3T MRI machines. % à reprendre
% This dataset contains multiple images such as a DSC 4D perfusion sequence, an injected T1 weighted sequence, a mask of the brain, glioma core, normal appearing white matter (NAWM) and three AIFs.

\subsubsection{Private dataset}\vspace{-0.25cm}
Additionally, DSC-MRI sequences acquired from 15 patients at Poitiers University Hospital were collected, including 8 LGG and 7 HGG patients confirmed by biopsy.
Imaging sequences were acquired with a 3T MRI machine (Skyra, Siemens Healthineers).
AIF extraction, tumor and normal-appearing white matter segmentations were performed by radiologists. 
% More details about the acquisiiton protocol are given in supplementary material. 
% BE CONSISTENT WITH BOTH DATA SETS. 
% A standard dose of gadoterate meglumine (0.1mmol/kg body weight; Dotarem, Guerbet, Villepinte, France) with a concentration of 0.5mmol/mL was administred. 
% An axial 2D gradient-echo EPI was employed with the following parameters: TR=1980ms, TE=30ms, flip angle=90°, FOV=220$\times$220mm², matrix=128$\times$128, voxel size=1.72$\times$1.72$\times$4.00mm$^3$, number of slices=30, the number of dynamic scans=75 and the time resolution=2.13s.

\subsubsection{Data split and preprocessing}\vspace{-0.25cm}
From the public dataset, 39 patients were assigned to the train set. 
All 25 remaining patients were assigned to the test set. 
All signals over time were extracted from the DSC-MRI images and transformed into concentration-time-curve  $C(t)$ with the following equation:
\begin{equation}
    C(t) = -\frac{1}{TE} \ln\left(\frac{S(t)}{S_0}\right)
\end{equation}
where TE is the echo time, $S(t)$ is the DSC signal over time and $S_0$ is the DSC signal baseline.
$C(t)$ and $C_a(t)$ (also named as AIF) were normalized between 0 and 1, by dividing all the curves by the maximum value found for each subject.
%Then, the same amount of $C(t)$ with a range of 0.1 were selected to be part of the training data to ensure signal intensity balancy.
%In total, 183,000 1D signals were extracted for the training dataset.

\subsection{Physics-informed autoencoder (PHAE)}

\begin{figure}[t!]
    \begin{minipage}[b]{1.0\linewidth}
      \centering
      \centerline{\includegraphics[width=5cm]{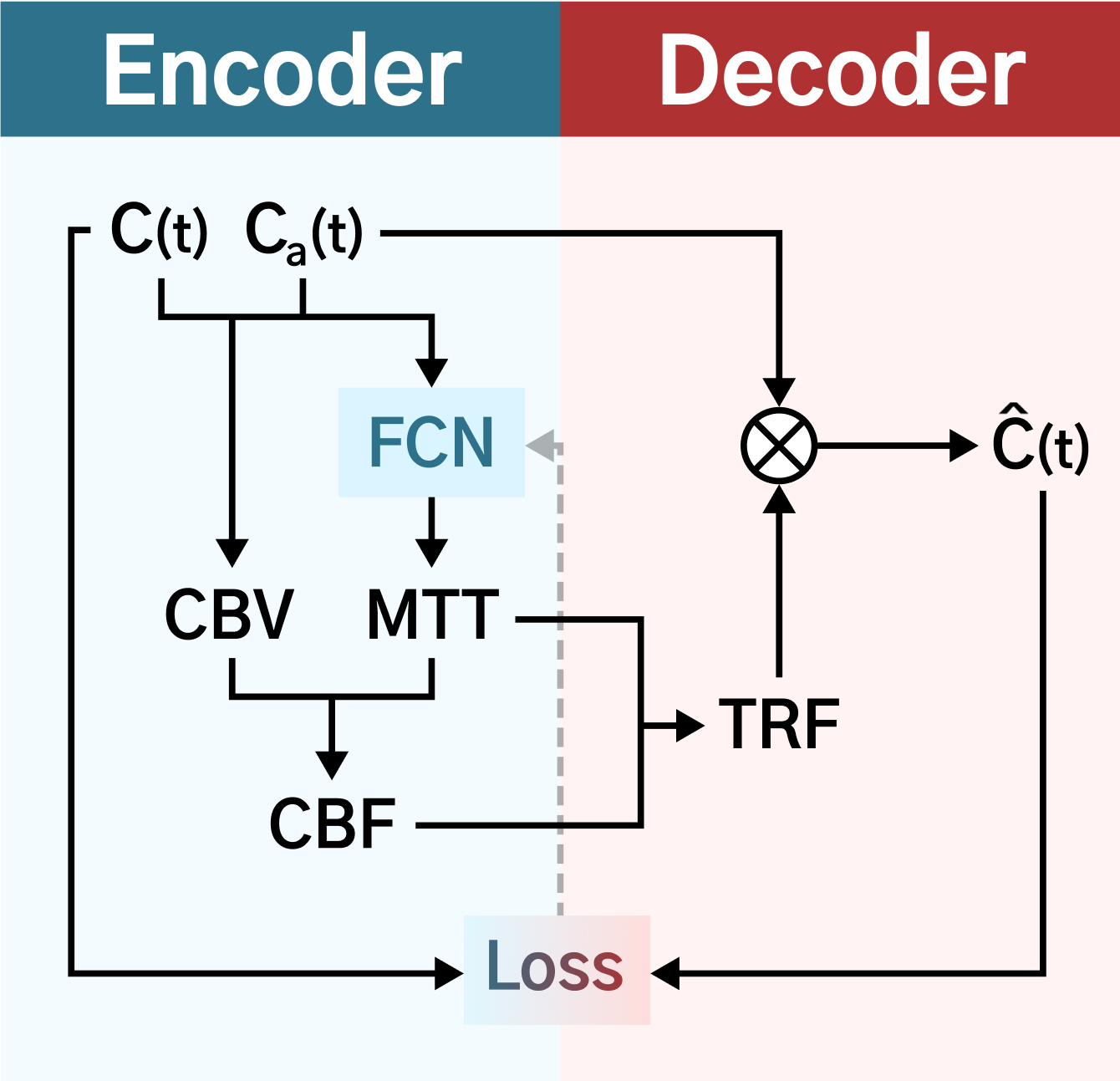}}
    \end{minipage}
    \caption{Workflow of the PHAE. $C(t)$: tissue concentration, $C_a(t)$: arterial concentration, $FCN$: Fully Convolutional Network, CBV: Cerebral Blood Volume, CBF: Cerebral Blood Flow, MTT: Mean Transit Time, TRF: Tissue Response Function. CBV is calculated from Eq.~\ref{eq:cbv}, CBF from Eq.~\ref{eq:cbf}, TRF from Eqs.~\ref{eq:rf}-\ref{eq:trf}, $\hat C(t)$ from Eq.~\ref{eq:conv}.}
    \label{fig:workflow}
\end{figure}

The proposed PHAE method is presented in Fig.~\ref{fig:workflow}.
This model is divided into an encoder that generates the perfusion parameters and a decoder that ensures the reliabilty of the perfusion parameters by reconstructing $C(t)$.

\subsubsection{Deep encoding network}\vspace{-0.25cm}

\begin{figure}[t!]
    \begin{minipage}[b]{1.0\linewidth}
      \centering
      \centerline{\includegraphics[width=7cm]{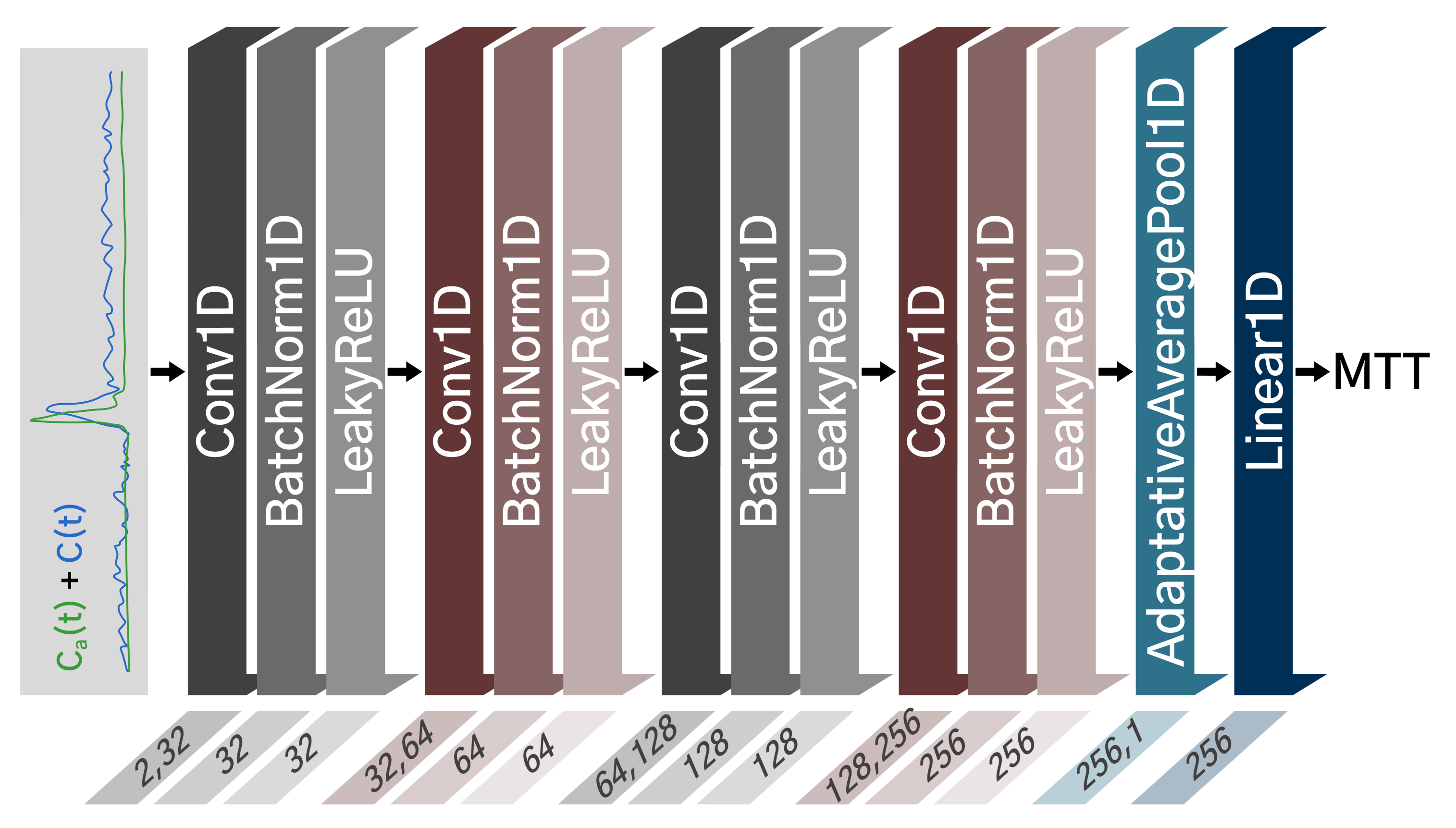}}
    \end{minipage}
    \caption{Architecture of the Fully Convolutional Network (FCN). Values represent the layer parameters.}
    \label{fig:pinn}
\end{figure}

\begin{table*}[!b]
    \small
    %Pour HAL
    %\scriptsize
    
    \centering
    \begin{tabular}{c|cccccc}
        Methods & AUC & Cut-off value & Sensitivity (\%) & Specificity (\%) & Accuracy (\%) & Inference time (s)\\
        \hline
        \rowcolor{LightCyan}
        oSVD~\cite{wu2003tracer} & 0.87 (0.69-0.99) & 1.77 (1.04-2.25) & 69.2 (50.0-100.0) & 100.0 (69.2-100.0) & 84.0 & 11.8 (7.09-17.33)\\
        % donner refs dans le tableau pour Tikhonv 
        Tikhonov~\cite{calamante2003quantification} & 0.88 (0.71-1.00) & 1.75 (1.01-1.95) & 69.2 (53.8-100.0) & 100.0 (75.0-100.0) & 84.0 & 33.6 (16.31-51.44)\\
        \rowcolor{LightCyan}
        % ALEX : je me suis permis de modifier ici pour faire apparaître "PHAE (ours)"
        PHAE (ours) & \textbf{0.90} (0.74-1.00) & 1.18 (1.05-1.40) & \textbf{76.9} (59.9-100.0) & 100.0 (84.6-100.0) & \textbf{88.0} & \textbf{8.4} (3.82-13.01)\\
        \hline
    \end{tabular}
    \caption{Diagnostic performance and computational time cost of CBF maps for differenciating LGG from HGG. Data in parenthesis represent the 95\% confidence intervals. The inference time represents the average time to process a single patient.}
    \label{tab:results}
\end{table*}

Fig.~\ref{fig:pinn} details the architecture of the proposed 1D Fully Convolutional Network (FCN). It takes as input a pair of $C(t)$ and $C_a(t)$. 
These inputs are fed into 4 blocks having each an 1D convolutional layer, an 1D batch normalization, a leaky ReLU with a negative slope of 0.02. 
The number of extracted feature maps is set to 32 and is multiplied by 2 for each following convolutional block.
The kernel size is set to 3 with a stride and a padding of 1 for each block. 
Then, an 1D average pooling is followed by a final linear layer with a single value as output corresponding to MTT. 
In the other hand, CBV is calculated by integrating $C(t)$ over $C_a(t)$ as followed:
\begin{equation}
    CBV = \frac{\int_0^\infty C(t) \;dt }{\int_0^\infty C_a(t) \; dt } \;\; \left[\frac{ml}{100g}\right]
    \label{eq:cbv}
\end{equation}
CBV are divided by the generated MTT to calculate CBF according to the central volume theorem%~\cite{stewart1893researches}:
\begin{equation}
    CBF = \frac{CBV}{MTT} \;\; \left[\frac{ml}{100g \cdot min}\right]
    \label{eq:cbf}
\end{equation}

%%%%%% CORRECTION
Here, MTT is the only perfusion parameters that is generated by the encoder as the CBV can be calculated beforehand, and the CBF can be computed through the previous equation to simplify the model.

\subsubsection{Physics-informed decoder}\vspace{-0.25cm}
To ensure the reliability of MTT values, a physics-informed decoder was developed based on the perfusion equations and using the previously generated perfusion parameters as input.
From MTT, a simulation of the residual function $R(t)$ is done using Lorentzian equation as proposed in~\cite{calamante2003quantification}:
\begin{equation}
    R(t) = \frac{1}{1 + \left(\frac{\pi \cdot t}{2 \cdot MTT} \right)^2}
    \label{eq:rf}
\end{equation}

Using a simulation for the generation of $R(t)$ is here to ensure a realistic shape without unwanted oscillations that could impact perfusion parameter estimate.
The residual function is multiplied by CBF to obtain TRF:
\begin{equation}
    TRF = R(t) \cdot CBF
    \label{eq:trf}
\end{equation}
Then, TRF is convolved with $C_a(t)$ to reconstruct a new $\hat C(t)$ according to the following equation:
\begin{equation}
    \hat C(t) = C_a(t) \otimes TRF
    \label{eq:conv}
\end{equation}

\subsubsection{Training details}\vspace{-0.25cm}
The more realistic the generated MTT values, the closer the $\hat C(t)$ reconstuctions match $C(t)$.
Therefore, the mean absolute error was used as a loss between $C(t)$ and $\hat C(t)$ to constrain the encoder in the possible MTT values to generate.
The encoder was trained during 65 epochs in approximatively 37 minutes.
The ADAM optimization algorithm was used with a learning rate of 0.0001 and without weight decay. 
The batch size was set to 1536.

\subsection{Experimental setup}

To generate perfusion maps for the test dataset, $C(t)$ and $C_a(t)$ of each patient were sent to the encoder only.
This results in a MTT value for each voxel of the brain, converting the 4D time series into a 3D image.
To evaluate the performance of our approach in generating MTT and thus CBF maps (via Eq.~\ref{eq:cbf}), two baseline deconvolution methods were used, oSVD~\cite{wu2003tracer} and Tikhonov~\cite{calamante2003quantification} as implemented in~\cite{fernandez2023perfusion}.

%\subsubsection{Glioma grading task}\vspace{-0.25cm}
By computing the mean CBF values of a lesion Region Of Interest (ROI) divided by the mean CBF values of a healthy ROI contralateral of the lesion, a ratio can be obtained.
This ratio can be used to determine glioma grading~\cite{hakyemez2005high}.
For that purpose, both tumor and normal-appearing white matter segmentations were used to calculate the ratio for each patient.

%\subsubsection{Performance metrics}\vspace{-0.25cm}
Receiver Operating Characteristic curve (ROC) analysis was used with Area Under the Curve (AUC) to quantify classification accuracy.
A cut-off ratio to distinguish LGG from HGG was then estimated for each method by maximizing both sensitivity and specificity with Youden's method~\cite{ruopp2008youden}.

To evaluate the robustness to noise, test dataset signals were progressively degraded from a SNR of 50 to 10.
$C_a(t)$ signals were systematically recomputed from the degraded DSC sequences.
The SNR was calculated as the mean of the DSC signal baseline divided by its standard deviation.
Gaussian noise was added to achieve specified SNR values.

%All analyses were conducted using Python. % with scikit-learn library~\cite{scikit-learn}.

\section{Results}

\subsection{Qualitative results}
\begin{figure}[t!]
    \begin{minipage}[b]{1.0\linewidth}
      \centering
      \centerline{\includegraphics[width=\textwidth]{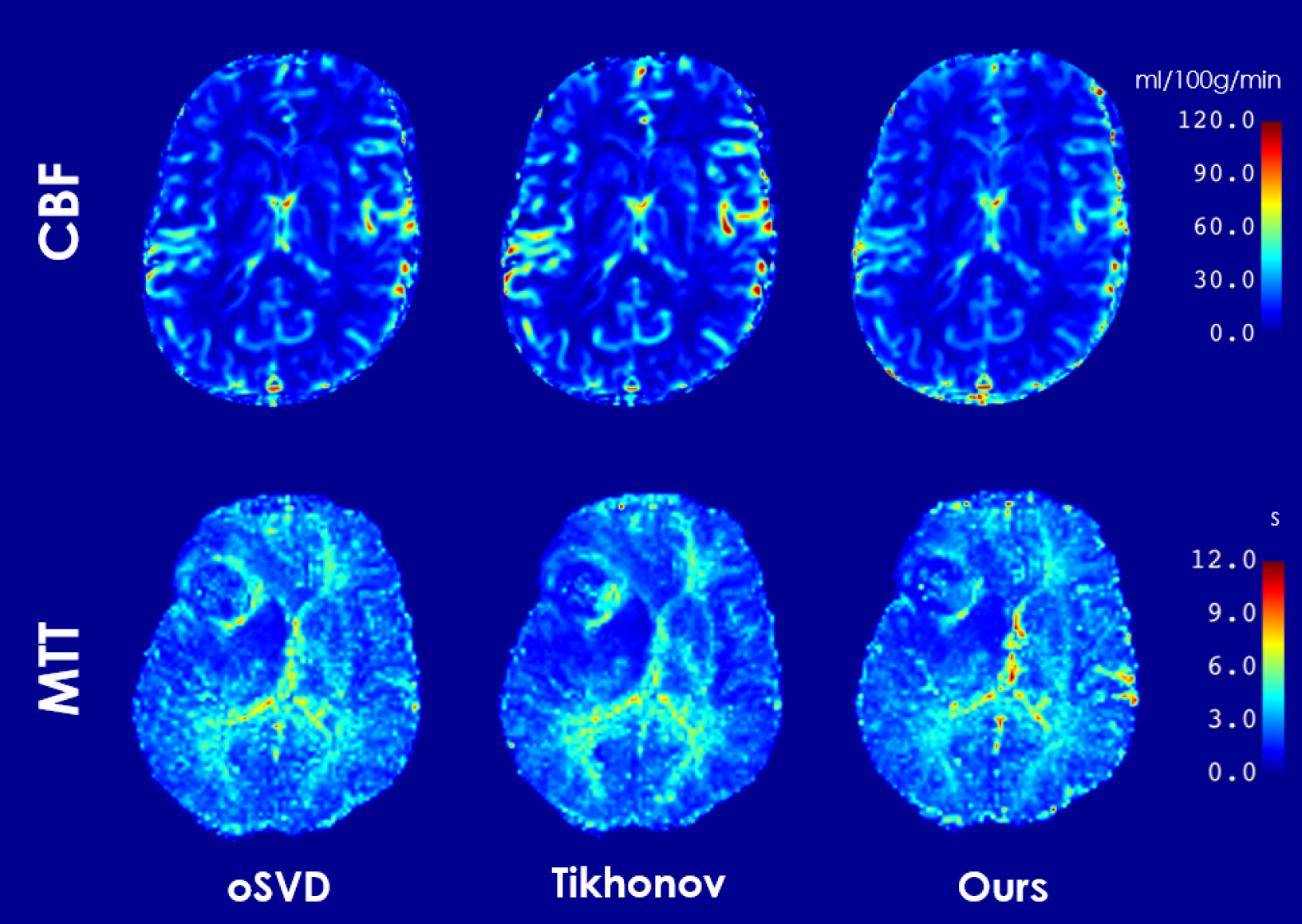}}
    \end{minipage}
    \caption{Comparison of CBF maps (grade 2 glioma) and MTT maps (grade 4 glioma) generated by oSVD~\cite{wu2003tracer}, Tikhonov~\cite{calamante2003quantification} and our methods.}
    \label{fig:cbf}
\end{figure}

The generated CBF and MTT maps are shown in Fig.~\ref{fig:cbf}.
The PHAE method maps are visually close from oSVD and Tikhonov methods.
The majority of variations in both CBF and MTT maps originate from gray matter and cerebrospinal fluid, with several outlier values observed across each method.
% EN FAIRE PLUS 

\subsection{Glioma grading performance}

\begin{figure}[t!]
    \begin{minipage}[b]{1.0\linewidth}
      \centering
      \centerline{\includegraphics[width=7.5cm]{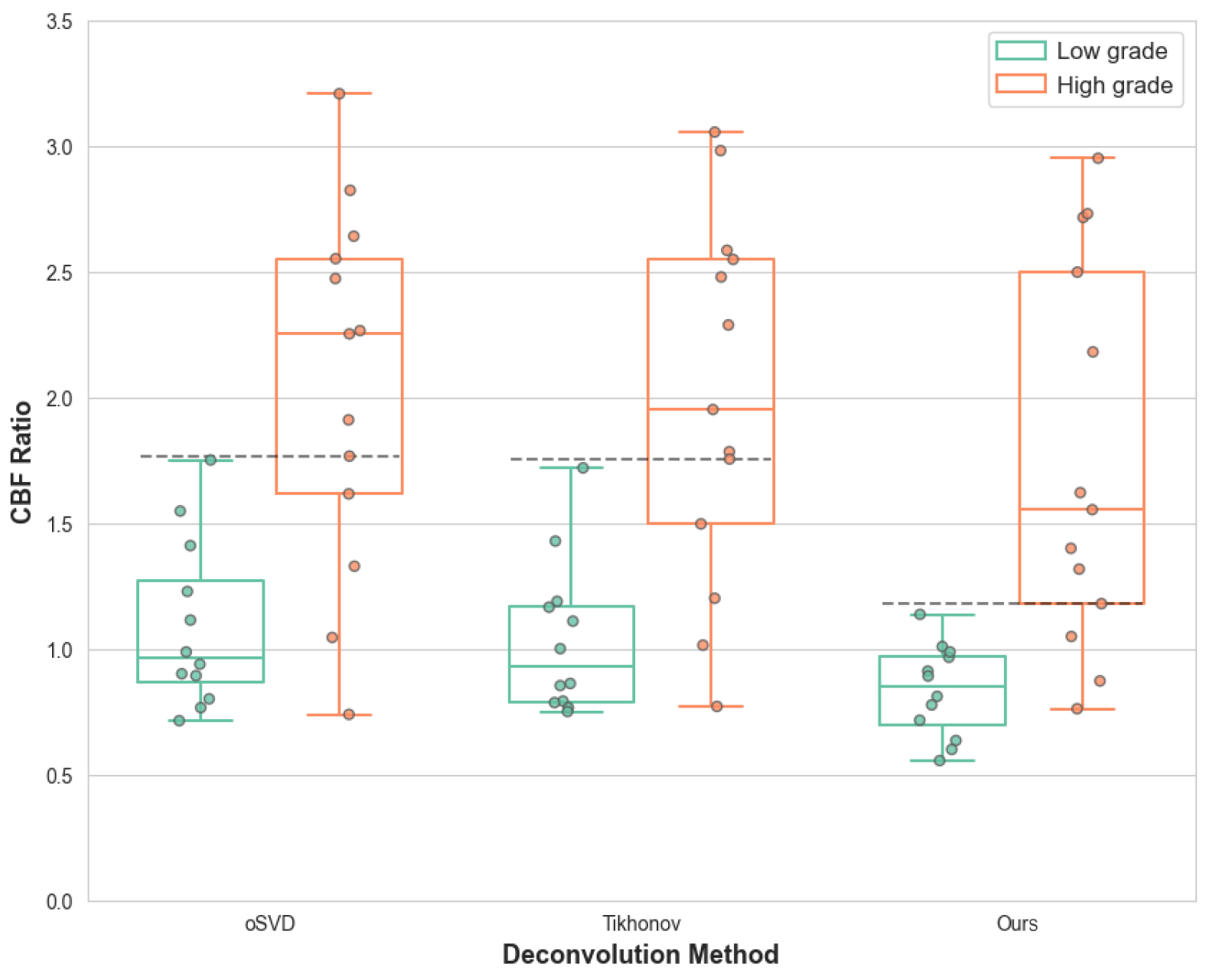}}
    %  \vspace{2.0cm}
    \end{minipage}
    \caption{Boxplots for each deconvolution methods based on CBF ratio to distinguish Low Grade Glioma (LGG) from High Grade Glioma (HGG). Each point represents a subject. The dashed lines represent the optimal cut-off values for the respective deconvolution methods.}
    \label{fig:boxplots}
\end{figure}

Results of deconvolution methods (oSVD, Tikhonov, and the PHAE) for generating the CBF maps are summarized in Table~\ref{tab:results}. 
% The different methods showed respectively cut-off values of 1.77 (1.04-2.25), 1.75 (1.01-1.95) and 1.18 (1.05-1.40).
The AUC were closely similar with respectively 0.87 (0.69-0.99), 0.88 (0.71-1.00) and 0.90 (0.74-1.00).
The accuracies for distinguishing LGG from HGG are 84\% (21 subjects correctly classified) for oSVD and Tikhonov, and 88\% (22 subjects correctly classified) for the proposed PHAE method.
% Sensitivity and specificity are respectively 69.2\% (50.0-100.0) and 100.0\% (69.2-100.0), 69.2\% (53.8-100.0) and 100\% (75.0-100.0), 76.9\% (59.9-100.0) and 100.0\% (84.6-100.0).
Fig.~\ref{fig:boxplots} shows boxplots categorized by glioma grading.
The PHAE method showed less spread CBF values for LGG.
The Mann-Whitney U test was employed to compare the distributions between LGG and HGG groups for each method. 
As a result, statistically significant difference was found, with p-values of 0.007, 0.003, and 0.0009 for oSVD, Tikhonov, and PHAE respectively.

\subsection{Robustness to noise evaluation}
\begin{figure}[h]
    \begin{minipage}[b]{1.0\linewidth}
      \centering
      \centerline{\includegraphics[width=8.5cm]{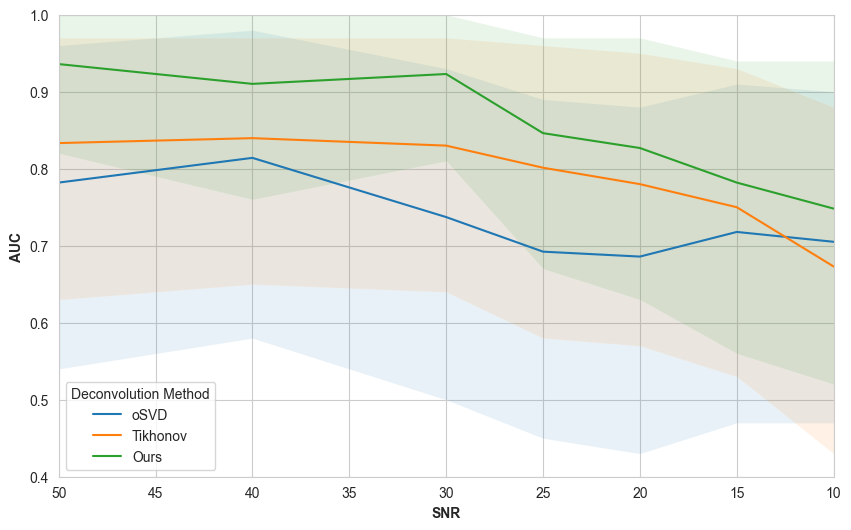}}
    %  \vspace{2.0cm}
    \end{minipage}
    \caption{AUC values with their confidence intervals as a function of SNR estimated for each deconvolution method.}
    \label{fig:degraded}
\end{figure}

After adding Gaussian noise, the ROC curves and AUC values were computed for each SNR level.
Fig.~\ref{fig:degraded} shows the AUC values of each deconvolution method to distinguish LGG from HGG as the SNR decreases.
As expected, artificially reducing the SNR tends to decrease the performance of each method.
The PHAE method outperforms oSVD and Tikhonov algorithms even under high noise levels (low SNR).

\subsection{Computational performance}
The inference time was measured (Intel Xeon Platinum 8253 CPU and 192GB RAM) for the generation of both MTT and CBF maps and were respectively 11.8 (7.09-17.33), 33.6 (16.31-51.44) and 8.4 (3.82-13.01) seconds on average per patient.

\comment{
\begin{figure}[t!]
    \begin{minipage}[b]{1.0\linewidth}
      \centering
      \centerline{\includegraphics[width=8cm]{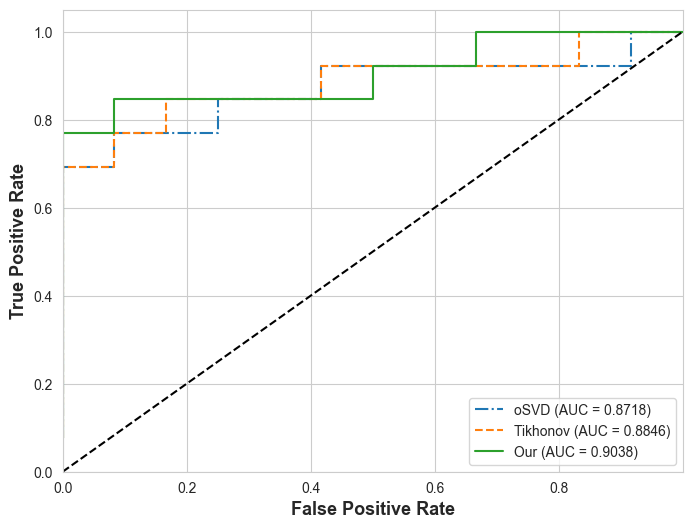}}
    %  \vspace{2.0cm}
    \end{minipage}
    \caption{Receiver operating characteristic curves for each deconvolution method to distinguish Low Grade Glioma (LGG) from High Grade Glioma (HGG) and their respective Area Under the Curve (AUC).}
    \label{fig:roc}
\end{figure}
}

\section{Discussion}
The proposed PHAE method demonstrated performance comparable to the oSVD and Tikhonov methods in differentiating LGG from HGG.
Interestingly, this confirms the feasibility of the proposed approach to learn to estimate the perfusion parameters without the need of third-party references.
Although CBF ratio cut-off threshold of our method are lower than others, it did not impact the quality of the resulting maps.

Among the test dataset, 4 patients were histologically diagnosed as grade 3 according to the WHO 2016 classification~\cite{louis20162016}.
All three methods failed to classify these glioma cases as HGG.
The tumor segmentations showed indeed lower CBF values, even though higher values would be expected.
Hypothetically, this could be explained either by an inaccurate segmentation of the active tumor core, or by a presence of LGG site in this segmentation, lowering CBF values.

The standard deconvolution methods require a longer processing time to generate the perfusion maps. To expedite this process, a compromise is often reached and leads to an increase estimating errors for the generated maps~\cite{asaduddin2024spinned}.
In contrast, the PHAE method can generate the perfusion maps in less than 9 seconds per subject, that is crucial for clinical applications to speed up diagnosis and preventive measures.

In the literature, multiple studies worked on simulating the residual function or TRF as proposed in~\cite{calamante2003quantification}.
In this work, the Lorentzian simulation function was chosen.
Interestingly, no significative change was found for mono-exponential, gamma function and Lorentzian function.
Other simulation functions such as bi-exponential, Fermi function or vascular model were not tested as they are not directly generated by MTT and result in a more complex model that was not the purpose of this work. 

Despite the reliable generation of both CBF and MTT, the DSC perfusion parameter T$_{max}$ was not generated.
Including this parameter in the PHAE may faciliate the generation of more accurate $\hat C(t)$ and will be one of the main focus of our future work. 

\section{Conclusion}
In summary, this paper presents a new approach for the perfusion parameter generation in DSC-MRI perfusion.
We proposed a physics-informed autoencoder that estimates reliable CBF and MTT in accordance with other standard deconvolution methods without any third-party perfusion maps used as reference.
The PHAE method showed better results for distinguishing LGG from HGG even in high presence of noise while needing less time to generate perfusion maps.


\section{Acknowledgments}
\label{sec:acknowledgments}
P.F., A.B., and N.D. are employees of Guerbet.
This work was supported by ANRT (CIFRE \#2023/1206).
%This work was granted access to the HPC resources of IDRIS under the allocation 2024-A0150314655 made by GENCI. 

\section{Compliance with Ethical Standards}
The private dataset comes from a study performed in accordance with the ethical standards as laid down in the 2013 Declaration of Helsinki.
All patients approved the use of their data with a written consent.


\bibliographystyle{IEEEbib}
% \bibliography{refs}

\end{document}